# Cooperativity in the enhanced piezoelectric response of polymer nanowires

*Luana Persano ‡ , Canan Dagdeviren‡, Claudio Maruccio‡, Laura De Lorenzis, and Dario Pisignano\**

Dr. L. Persano, Prof. D. Pisignano
National Nanotechnology Laboratory of Istituto Nanoscienze-CNR, via Arnesano, I-73100 Lecce (Italy)
Dr. C. Dagdeviren
Department of Materials Science and Engineering, Frederick Seitz Materials Research Laboratory, and Beckman Institute for Advanced Science, University of Illinois at Urbana-Champaign, Urbana, IL 61801, USA.
Dr. C. Maruccio
Dipartimento di Ingegneria dell'Innovazione, Università del Salento, via Arnesano I-73100 Lecce (Italy).
Prof. Laura De Lorenzis
Institut für Angewandte Mechanik, Technische Universität Braunschweig, Bienroder Weg 87 Campus Nord, D-38106 Braunschweig (Germany).
Prof. D. Pisignano
Dipartimento di Matematica e Fisica "Ennio De Giorgi", Università del Salento, via Arnesano I-73100 Lecce (Italy)
E-mail: dario.pisignano@unisalento.it

‡ These authors contributed equally to this work.

Piezo-electricity, a Greek term for pressure-induced electricity, is the capability of a material to polarize by means of spatially-separated electrical charges of opposite sign, in response to an external stress that produces a mechanical deformation. Generally, charges accumulate at two opposite side surfaces of the material body, and in absence of short-circuited contacts, a voltage bias is generated. This effect can be observed in materials whose crystalline state has no center of symmetry (so-called non-centrosymmetric), including polymers and biological systems.[1] To date, piezoelectricity represents one of the most valuable alternative source of energy with an associated fast-growing investment market and potential applications spanning across a wide range of fields, such as information and communications, industrial automation, healthcare and medical monitoring, defense industry, automation and robotics.[2] Indeed, the capability to harvest energy from small mechanical forces, through pressure, vibration, bending, elongation and compression, is today subject of extensive research on both materials and device geometries, and the related development of self-powered wireless devices could be of great importance for the internet of things, that is for inter-connecting individual uniquely identifiable objects and bodies.[3, 4] In this respect, piezoelectric micro and nanostructures have demonstrated improved properties that enable new functionalities not achievable with their bulk counterpart. Most of these are related to reduced dislocations and superior mechanical properties.[5-7] For instance, in pioneering works by the Wang group aligned arrays and multilayer stacks of zinc-oxide and lead zirconate titanate nanowires have been exploited to power light-emitting and wireless devices.[8, 9]

In this framework, piezoelectric polymers are very promising, since they can also provide structural flexibility and toughness, as well as low cost, improved biocompatibility, and ease of processing. In particular, the device-integration of Polyvinylidenefluoride (PVDF) and its copolymers is attracting increasing interest,[10-12] because their micro and nanostructures such as films, belts and fibers have shown unique advantages in terms of material functionality and piezoelectric response, and self-poling during nanofabrication.[13-17] Electrospinning is especially effective in this respect, producing self-poled piezoelectric nanofibers due to the very high stretching forces exerted on electrified solution jets.[17] Consequently, polymer molecules mainly align parallel to the fiber longitudinal axis,[18] and piezoelectric material phases are favored compared to films.[16, 19] Furthermore, aligned arrays of PVDF-based fibers generally exhibit still superior piezoelectric performances.[20-22] Most often, these fibers are aligned with low density and provide sub-monolayer coverage of solid supports, namely they are separated by distances of the order of microns from their nearest neighbors in deposited strands. This configuration results in open-circuit currents which correspond to the sum of currents generated by each single nanowire in the generator.[23] Dense ($10^7$ fibers/mm$^2$) arrays of electrospun aligned nanofibers of poly(vinylidenefluoride-*co*-trifluoroethylene) [P(VDF-TrFe)] offer exceptional piezoelectric characteristics and output *voltage* significantly enhanced with respect to individual fibers.[19] Such arrangement is characterized by large sensitive areas (tens of cm$^2$) and light weight, and it may be bent or twisted without fracture. However, the improved voltage output from aligned arrays of polymer piezoelectric nanostructures in mutual contact cannot be explained by conventional circuit theory. The in-depth understanding of this mechanism and the assessment of its possible general validity for nanofibers regardless of their constituent material or fabrication process would be very important for realizing improved mechanical energy-harvesting architectures.

Here, we provide a detailed insight into piezoelectric energy generation from arrays of polymer nanofibers. For sake of comparison, we firstly measure individual P(VDF-TrFe) fibers at well-defined levels of compressive stress. Under an applied load of 2 mN, single nanostructures generate a voltage of 0.45 mV. We show that under the same load conditions, fibers in dense arrays exhibit a voltage output higher by about two orders of magnitude.

Numerical modelling studies demonstrate that the enhancement of the piezoelectric response is a general phenomenon associated to the electromechanical interaction among adjacent fibers, namely a cooperative effect depending on specific geometrical parameters. This establishes new design rules for next piezoelectric nano-generators and sensors.

P(VDF-TrFe) fibers were electrospun by a potential of 30 kV onto a collector disk with sub-cm width rotating at 4,000 rpm (see Experimental Section). Strands of mutually isolated fibers were deposited onto glass coverslips, mounted on the rotating collector. Dense arrays of fibers were directly spun onto the collector surface. Representative scanning electron microscope (SEM) micrographs are displayed in Figure 1a and 1b. Fibers are smooth, with uniform diameter over their length. Array samples appear as in the photograph in the inset of Fig. 1c.

To evaluate the piezoelectric response of fibers under specific compressive loads, flexible and thin Cu wires with a layer of silver epoxy were used to define contacts. For electrical measurements, both isolated fibers and arrays were positioned onto glass substrates, and a triboindenter TI 950 (Hysistron) equipped with a flat-ended cylinder sapphire tip (1 mm diameter) was used to apply calibrated forces. A scheme of the set-up used is reported in Figure 1c. Under applied forces of 0.8-2.0 mN, fiber displacements were in the range of 30-100 nm, related to the viscoelastic properties of the material and to the fibrous shape. No significant difference was appreciable in the displacements of individual fibers and arrays under the same applied force. A lock-in amplifier was used to collect the open-circuit voltage from the fibers when a well-defined level of compressive force is delivered. Data in Figure 1d highlight linear variations of the output voltage with applied force, presenting slopes of 0.2 V/N and 15 V/N for single fibers (green dots) and arrays (red dots), respectively. The piezoelectric response of P(VDF-TrFe) fibers in the arrays is strongly enhanced with respect to isolated fibers. Additional information evidencing the different response in the two systems were collected by dynamic loading-unloading cycles (Figure 1e and 1f). During compression, these measurements showed a well-behaved, periodic alternation of negative and positive peaks of the open-circuit voltage outputs, corresponding to the application and release of the stress, respectively. For identical applied loads in the range 0.8-2.0 mN, piezo-voltages for isolated fibers and arrays were peaked at 0.15-0.45 mV and 10-30 mV, respectively. Hence, it is clear that dense arrays of aligned piezoelectric polymer fibers yield enhanced voltage response under identical applied forces and consequent strain conditions.

To explain in depth this behavior, we developed extensive numerical simulations through a finite element multiphysics environment, describing the complex electromechanical interaction among fibers at microscale which affects the resulting polarization measured along the direction ($x_3$) of the fiber length, $L$ (Fig. 2a). The piezoelectric polymer is described by the constitutive equations of linear piezoelasticity (see Supporting Information), and the mutual interaction of fibers is mainly due to contact between adjacent elements as a result of the applied loading. Compressive forces are applied by means of pressing plates in the direction ($x_2$) orthogonal to the plane defined by the array of fibers and identified by axes $x_1$ and $x_3$ as schematized in Fig. 2a, and the consequent voltage distributions due to the applied pressure are then determined. For instance, the component of the electric displacement along the fiber length is $D_3 = \sum_{j=1}^{3} d_{3j} \sigma_{jj} + k_{33} E_3$, where $d_{3j}$ the are piezoelectric coefficients, $\sigma_{jj}$ are the stress components, $k_{33}$ is the dielectric permittivity coefficient and $E_3$ is the electric field component in $x_3$ direction. The output voltage bias, $V_{out}$, at the two ends of fibers is then obtained directly from such analysis.

The inter-fiber interaction is found to have different effects depending on the stacking directions of individual nanostructures ($x_1$ or $x_2$ as shown in Fig. 2b), and on the fiber cross-sectional shape. Both circular cross-sections with radius, $R$, well-describing our electrospun fibers, and rectangular cross-sections of width, *W*, and thickness, *T*, are considered. Fibers with rectangular cross-section are a useful basis of comparison, since modeling their stacks with zero inter-fiber spacing leads to describe bulk piezoelectric samples. Full details on numerical modeling are reported in the Supporting Information. In the following, we firstly analyze the behavior of single piezoelectric fibers, and then analyze how the output voltage is affected by the electromechanical interaction of adjacent fibers.

*Single piezoelectric fibers.* For fibers of both cylindrical and rectangular shape, increasing the length by a given factor leads to an increase of the output voltage by roughly the same factor as shown in Figure 3a, for same values of applied pressures. This allows the nanogenerator to be described by modeling segments of fibers much shorter than in experiments, thus significantly saving computational time. Validating such $V_{out}(L)$ dependence, one legitimates analyzing fibers of shorter length, and using results to compute those for any desired length through linear correlation. In addition, for fibers with rectangular cross-sections we found that $V_{out}$ does not depend on the width, *W*, under a constant pressure, whereas it is enhanced upon increasing the thickness, *T*, since the system correspondingly becomes less stiff.

A significant improvement of the piezoresponse is obtained by using cylindrical fibers. Indeed, under the same applied force, the output voltage from a cylindrical fiber is around 2.2 times that from a fiber with rectangular cross-section (Fig. 3b). To further highlight the role of the cross-sectional geometry, several distortion factors are applied to the circular shape as illustrated in Fig. 3c. In fact, the piezoelectric response of fibers with elliptical cross-section is enhanced upon increasing the *b/a* ratio, where *a* and *b* indicate the ellipse axis along the $x_1$ and the $x_2$ direction, respectively. This establishes new design rules for electrospun piezoelectric nanofibers, in which an elliptical cross-section can be obtained as a result of skin collapse following rapid solvent evaporation from the surface of electrified jets, of partial flattening when fibers impact on the collecting surface upon electrospinning with a slowly evaporating solvent component, or of both.[19, 24]

*Cooperative effects in arrays of aligned piezoelectric polymer fibers*. No increase of $V_{out}$ is found upon placing many fibers with rectangular cross-section in mutual contact and in parallel under a constant pressure, i.e. building an array along the $x_1$ (planar) direction which corresponds to considering a continuous, bulky piezoelectric film of thickness $T$. On the contrary, a remarkable enhancement of the piezoresponse is observed by increasing the number of adjacent cylindrical fibers along $x_1$. The comparison, performed for the same pressing geometry and boundary conditions, is displayed in Figure 4a. An increment of $V_{out}$ up to three times is achieved by a monolayer of aligned cylindrical fibers in mutual contact compared to a film of same thickness and under the same applied pressure. Such effect is mainly due to the electromechanical interaction among a few adjacent cylindrical fibers, giving rise to a *cooperative effect in the plane* of the array that restrains the transverse deformation and correspondingly increases the transverse stresses as shown in Fig. 4b. This mechanism in turn leads to an increase of the piezoelectric response along the longitudinal fiber axis, and takes place when the array starts to be built. In fact, when the number of parallel fibers in the monolayer exceeds five, no further increase of $V_{out}$ is observed.

A more complex cooperativity is found *along the out-of-plane* ($x_2$) direction, and obtained by stacking several layers of piezoelectric fibers. For a given value of the force applied to pressing plates and for identical pressing geometry and boundary conditions, the presence of more than a layer of fibers aligned along the vertical direction enhances the overall piezoelectric response of the system up to two orders of magnitude with respect to a monolayer, as shown in Figure 5a. Both the reduction of the mechanical stiffness along the thickness direction and the inter-fiber electromechanical contact interactions concur to such enhancement. These two effects can be assessed independently by comparing the results for the out-of-plane stacked cylindrical fibers and for a bulk system having the same total thickness. For a large number of stacked cylindrical fibers, the ratio of the output voltage to that of an equally thick piezoelectric bulk body stabilizes around a factor of two as shown in Fig. 5a. These findings indicate that the thickness-related reduced stiffness of the array yields a piezoresponse enhancement which increases roughly linearly upon increasing the number of stacked layers (as for black dots in Fig. 5a), whereas a further doubling of the piezoresponse is to be attributed to the electromechanical contact interaction of cylindrical fibers in the array constituting the nanogenerator. The corresponding transverse stresses in the fiber arrays are displayed in Fig. 5b. Through computational homogenization, we find that the cooperative behavior would correspond to the following, asymptotic effective piezoelectric coefficients: $\bar{d}_{31}$ =19.6 pC/N, $\bar{d}_{33}$ =-29.3 pC/N, $\bar{c}_{11}$ =1.5 GPa and $\bar{k}_{33}$ =10.1 $k_0$ where $k_0$ is the vacuum permittivity (Supporting Information). Consequently the macroscopic effective piezoelectric voltage constants $\bar{g}_{31}$ and $\bar{g}_{33}$ would be as high as 219.7×10$^{-3}$ Vm/N and -329.6×10$^{-3}$ Vm/N, respectively. Such description agrees well with experimental results for the different values of investigated applied forces, for both individual piezoelectric nanofibers and for their arrays, as reported in Table 1. Discrepancies between measured data and model predictions can be explained by considering the slight experimental dishomogeneity in fiber orientation, geometry and distribution in the array.

In summary, cooperative effects in the response of piezoelectric fibers due to electromechanical interactions at the microscale were predicted. The effect of geometry variables on the output of nanogenerators and strain sensors was investigated in detail through single-fiber experiments and parametric modeling studies, evidencing the enhancement of piezoelectric performances of cylindrical nanofibers and of their arrays compared to bulk architectures. Output piezo-voltages enhanced by two orders of magnitude are achieved by arrays of uniaxially aligned fibers due to cooperative electromechanical effects. The analysis carried out in this work is of general applicability for fibers featuring linearly elastic behavior within a small deformation regime. Also, finite deformations would likely introduce quantitative but not qualitative changes to the observed effects. Examples of other piezoelectric fibers whose response can be enhanced by cooperative effects in arrays include those made of liquid crystalline polymers such as poly(γ-benzyl α- L-glutamate),[25] biomaterials like β-glycine,[26] and composites with ceramic particles.[27] A variety of fields may be envisaged, where these findings can find application during device design and realization, including vibration sensing, power sources,[28] and especially self-powered and wearable electronics,[29] smart textiles and stick-on biomedical patches for health monitoring.[30]

*Experimental Section*

*Nanofiber device fabrication.* P(VDF-TrFe) was purchased from Solvay Solexis and dissolved in 3:2 volume ratio of dimethylformamide (DMF)/acetone (Sigma Aldrich). Electrospinning (ES) was performed by placing the polymer solution into a plastic syringe tipped with a 27-gauge stainless steel needle. Voltage bias were applied to the metal needle from a high voltage supply (EL60R0.6-22, Glassman High Voltage). During ES, the injection flow rate was kept constant at 1 mL h$^{-1}$ with a syringe pump (Harvard Apparatus). A grounded cylindrical collector (diameter = 8 cm), was placed at a distance of 6 cm from the needle. Strands of isolated fibers were deposited onto borosilicate glass coverslips directly mounted on the rotating collector, while dense arrays of fibers were directly deposited on the surface of the collector. Produced fibers are a few centimeters long, limited by the dimensions of the glass slides used. The morphological analysis was performed by SEM with a Nova NanoSEM 450 system (FEI), using an acceleration voltage around 5 kV and an aperture size of 30 μm.

*Indentation and voltage measurements.* A triboindenter TI 950 (Hysistron) equipped with a flat-ended cylinder sapphire tip (1 mm diameter) was used to apply calibrated forces. A custom data-recording system consisting of a lock-in amplifier (SR830, Standard Research Systems), a multiplexer (FixYourBoard.com, U802), and a laptop was used to capture open-circuit voltage data. Flexible thin Cu wires with a layer of silver epoxy (Ted Pella, Inc) were used to connect the terminations of the fiber arrays. The piezoelectric measurements were carried out at room temperature.

*Numerical simulations.* To numerically model the cooperative electromechanical behavior of nanostructures, a finite element multiphysics simulation environment was developed. Fibers were discretized with linear 8-node brick elements and the interaction between individual wires is due to contact between adjacent fibers as a result of the applied loading. The implemented formulation is based on the classical master-slave concept. A detailed description of the defined model is reported in the Supporting Information.

**Acknowledgements**
L. P. and C. D. thank Prof. J. A. Rogers for helpful discussion and continued support. The research leading to these results has received funding from the European Research Council under the European Union's Seventh Framework Programme (FP/2007-2013)/ERC Grant Agreements n. 306357 (ERC Starting Grant "NANO-JETS") and n. 279439 (ERC Starting Grant "INTERFACES"). C.M. acknowledges the support from the Italian MIUR through the project FIRB Futuro in Ricerca 2010 "Structural mechanics models for renewable energy applications" (RBFR107AKG).

Figures and Tables

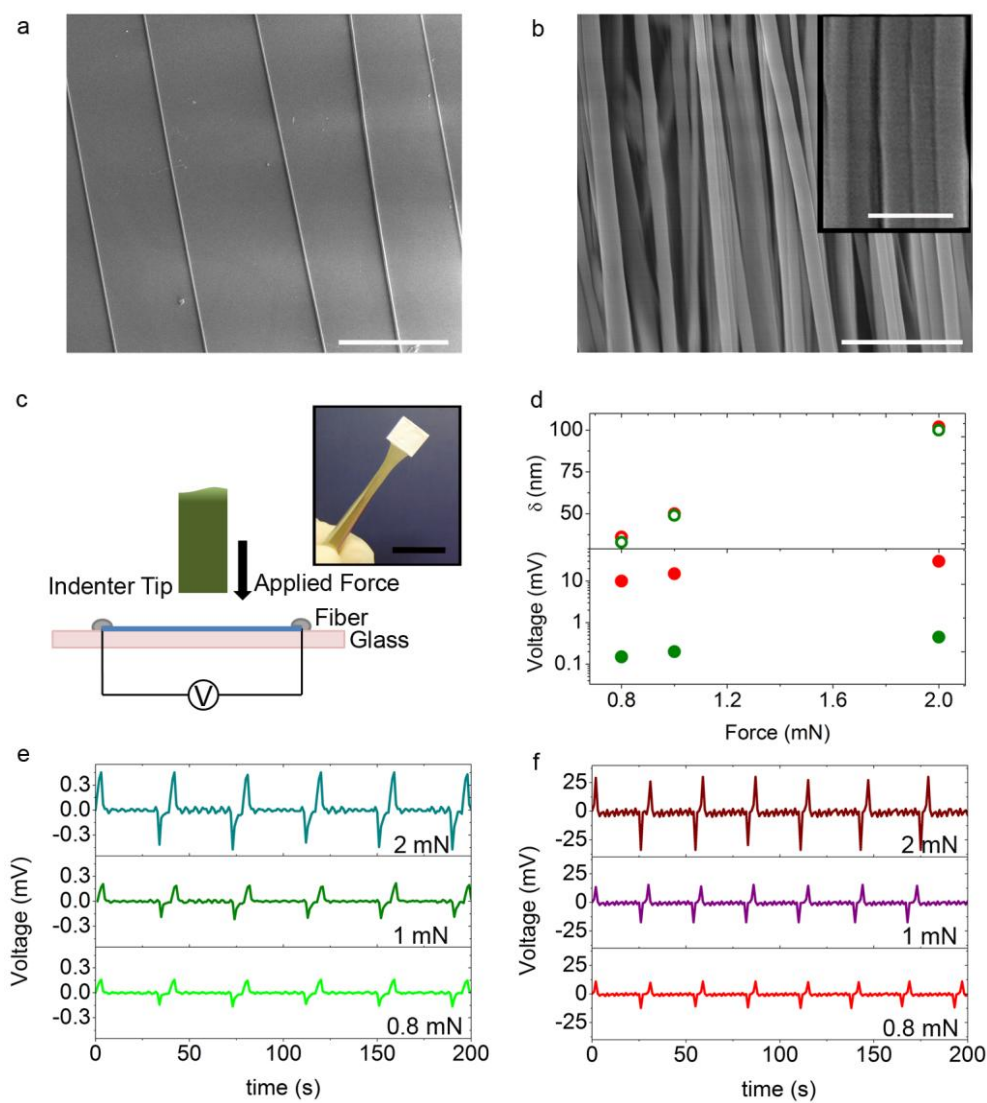

**Figure 1.** SEM micrograph of (a) a strand of mutually isolated fibers (scale bar, 20 μm) and (b) a dense array of aligned fibers at different magnification (scale bar, 3 μm). Inset: magnification of aligned fibers with line mutual contact along their side (scale bar, 1 μm). (c) Schematic illustration of the experimental setup for force-indentation measurements. Inset: photograph of a fiber array sample. (d) Measured displacement, δ, and voltage response of single fiber (green dots) and array of fibers (red dots) at different applied forces. Measured output voltage under repeated load/unload cycles for single fiber (e) and array of fibers (f). From top to bottom panels, applied loads are 2.0 mN, 1.0 mN and 0.8 mN, respectively.

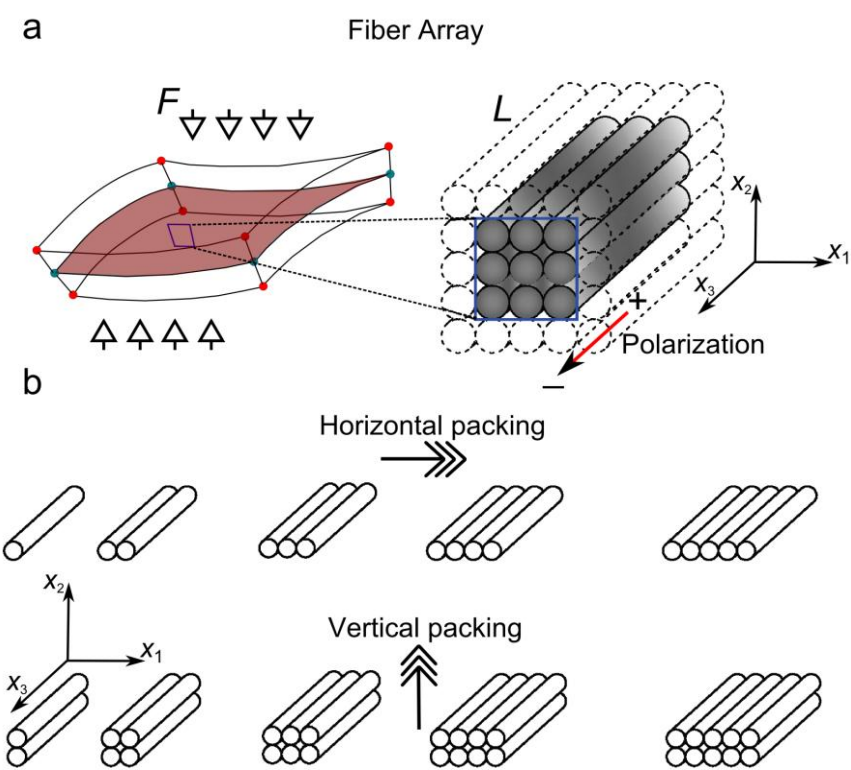

**Figure 2.** (a) Scheme of the PVDF-based fiber array structure at macro- and micro-scale. *F*: applied compressive force. *L*: fiber length. (b) Packing of fibers in the vertical and horizontal directions.

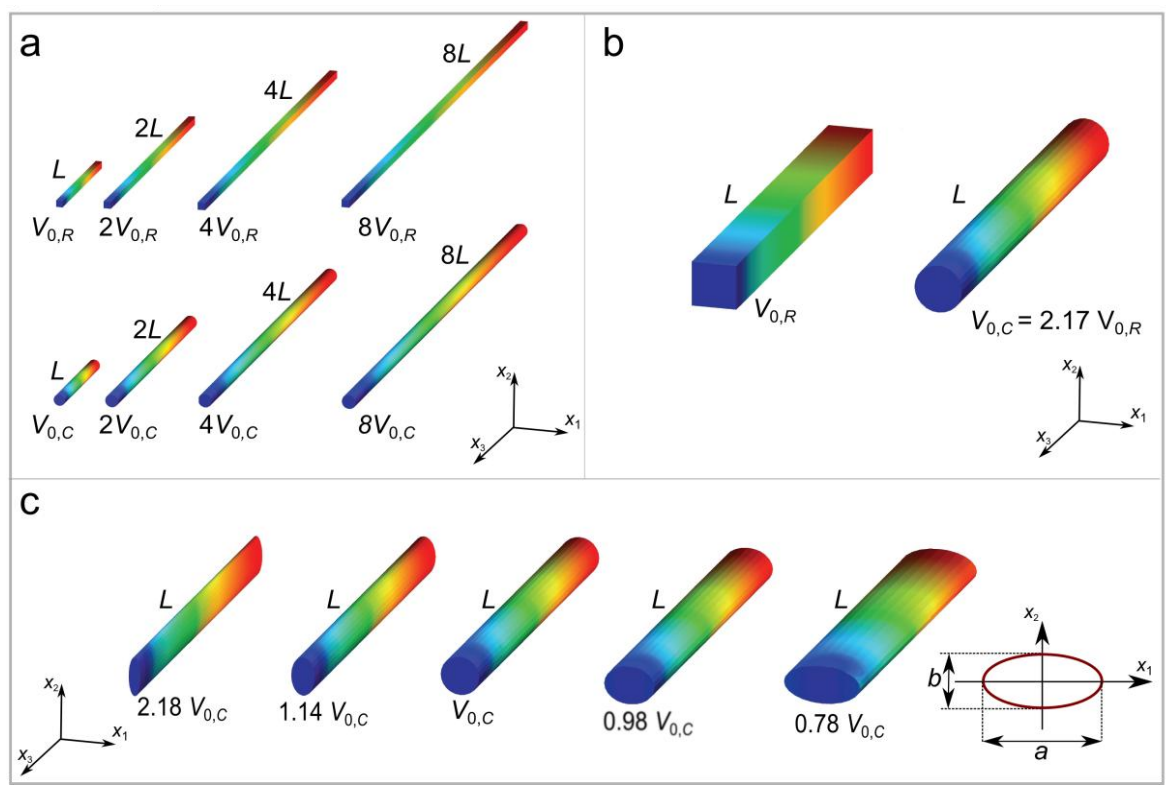

**Figure 3.** (a) Voltage distribution on the surface of nanofibers with rectangular (*R*, top row) and circular cross-section (*C*, bottom row), and having different length (*L*, 2*L*, 4*L*, 8*L*). Same pressures (120 Pa) are applied in all the investigated cases. For each fiber, red (blue) corresponds to high (low) voltage values, and the overall output voltage bias at termination is reported, highlighting a clear proportionality between fiber length and generated voltage. $V_{0,R}$ = 0.90 μV and $V_{0,C}$ = 1.95 μV indicate values obtained with a fiber of length *L*. (b) Comparison of nanofibers with same length, *L*, and either rectangular or circular cross-section. (c) Comparison of nanofibers with elliptical cross-section and different *b/a* ratios. For each fiber, we show the corresponding output voltage which range from 2.18 to 0.78 times $V_{0,C}$ upon varying the *b/a* ratio. Bottom inset: ellipse cross-section and *a* and *b* axes.

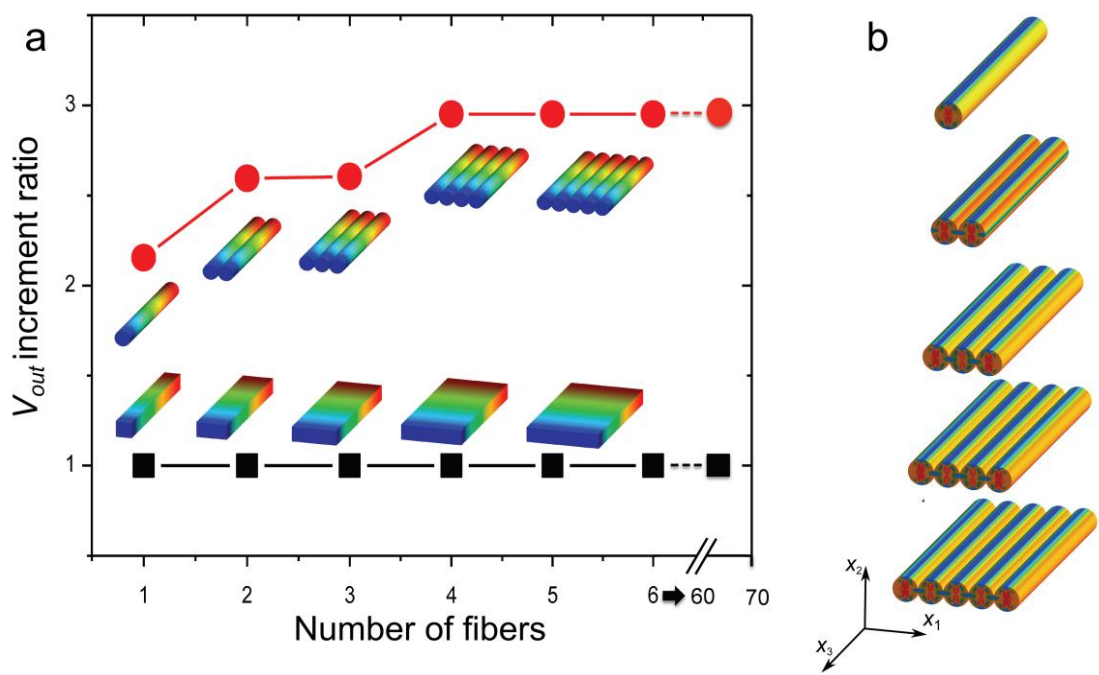

**Figure 4.** Cooperativity for nanofibers with different cross-sectional shape. (a) Dependence of output voltage on the number of aligned fibers. Cooperativity is found only in cylindrical fibers, leading to an increment of $V_{out}$ up to three times compared to single fibers. The insets show voltage maps. Applied pressure = 120 Pa. (b) Corresponding contour levels of stress ($\sigma_1$) distribution in the nanofibers. Maximum and minimum stress values, corresponding to the used color scales (i.e. to red and blue, respectively) are, from top to bottom: +0.098 and −0.368 MPa, +0.117 and −0.489 MPa, +0.122 and −0.507 MPa, +0.156 and −0.651 MPa, +0.166 and −0.694 MPa.

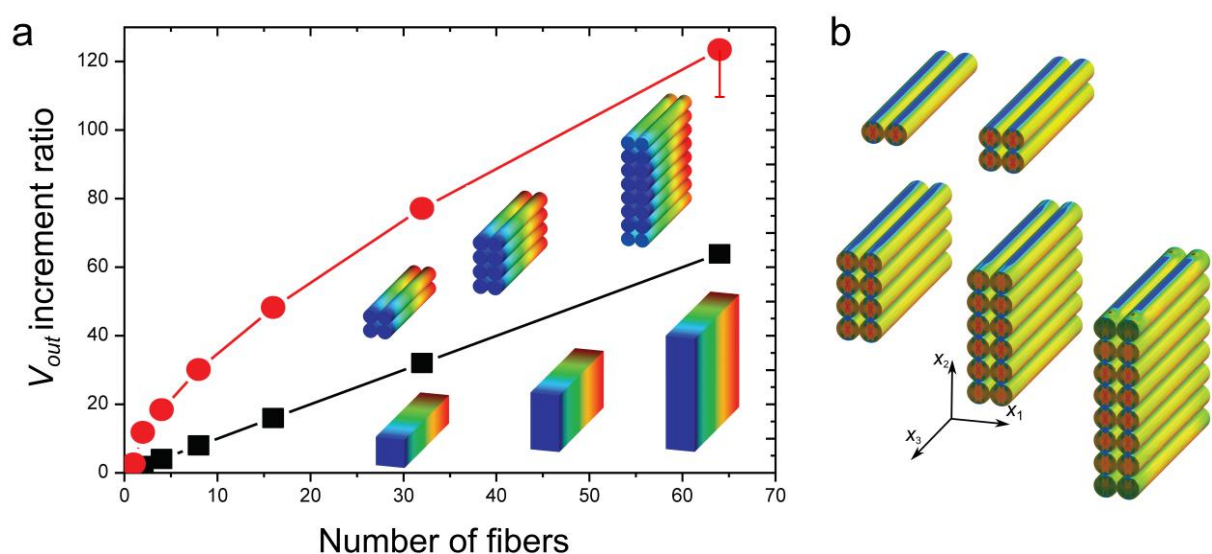

**Figure 5.** (a) Dependence of output voltage on the number of fiber layers stacked in the $x_2$ direction. The overall sample thickness for 64 fibers is about 25 μm. The insets show voltage maps. Applied pressure = 120 Pa. Maximum error bar indicative of the decrease of $V_{out}$ estimated on the basis of the study of a possible misalignment of ±10° among fibers as reported in the Supporting Information. (b) Corresponding contour levels of stress ($\sigma_1$) distribution in the nanofibers. Maximum and minimum stress values, corresponding to the used color scales (i.e. to red and blue, respectively) are, from top-left to bottom-right: +0.117 and −0.489 MPa, +0.163 and −0.656 MPa, +0.186 and −0.730 MPa, +0.195 and − 0.876 MPa, +0.231 and − 0.952 MPa.

|  | Single fiber $V_{out}$ (mV) |  | Array $V_{out}$ (mV) |  |
|---|---|---|---|---|
| F (mN) | Experiment (~) | Model | Experiment (~) | Model |
| 0.8 | 0.15 | 0.20 | 10 | 13.8 |
| 1.0 | 0.20 | 0.25 | 15 | 17.2 |
| 2.0 | 0.44 | 0.49 | 30 | 34.4 |

**Table 1.** Comparison between experimental and numerical results.

# Cooperativity in the enhanced piezoelectric response of polymer nanowires

*Luana Persano‡, Canan Dagdeviren‡, Claudio Maruccio‡, Laura De Lorenzis, and Dario Pisignano\**

Dr. L. Persano, Prof. D. Pisignano
National Nanotechnology Laboratory of Istituto Nanoscienze-CNR, via Arnesano, I-73100 Lecce (Italy)
Dr. C. Dagdeviren
Department of Materials Science and Engineering, Frederick Seitz Materials Research Laboratory, and Beckman Institute for Advanced Science, University of Illinois at Urbana-Champaign, Urbana, IL 61801, USA.
Dr. C. Maruccio
Dipartimento di Ingegneria dell'Innovazione, Università del Salento, via Arnesano I-73100 Lecce (Italy).
Prof. Laura De Lorenzis
Institut für Angewandte Mechanik, Technische Universität Braunschweig, Bienroder Weg 87 Campus Nord, D-38106 Braunschweig (Germany).
Prof. D. Pisignano
Dipartimento di Matematica e Fisica "Ennio De Giorgi", Università del Salento, via Arnesano I-73100 Lecce (Italy)
E-mail: dario.pisignano@unisalento.it

‡ These authors contributed equally to this work.

**SUPPORTING INFORMATION**

## 1. Piezoelectric polymer fiber nanogenerator scheme

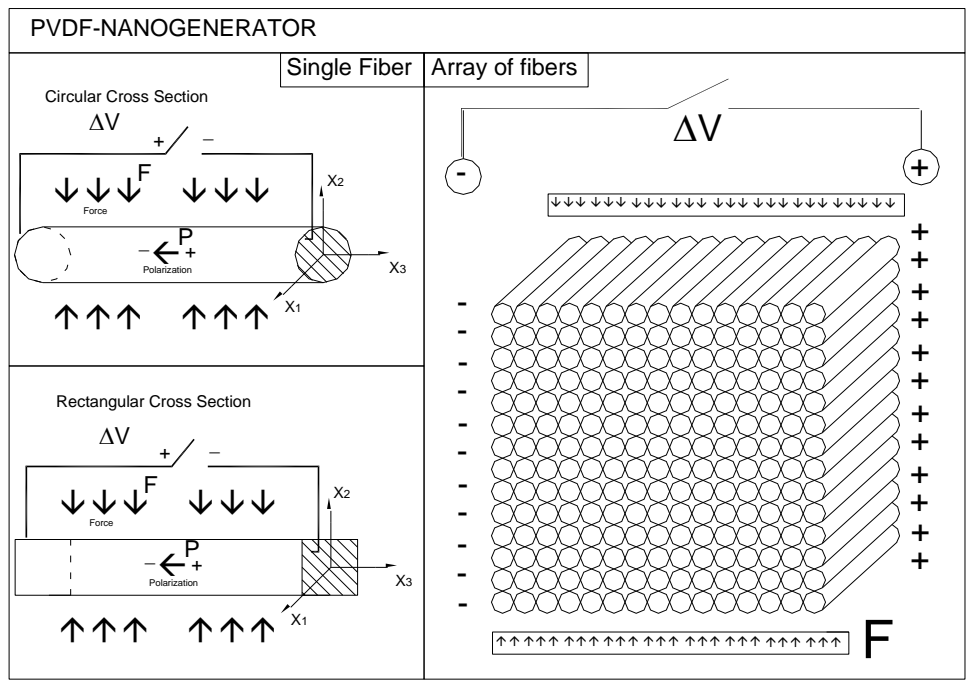

Fig. S1. Schematic view of the studied piezoelectric nano-generator: single fiber (square and circular cross-section) and array of fibers.

## 2. Linear piezoelasticity and constitutive equations

In piezoelectric phenomena an electric potential gradient causes a material deformation, and vice versa. The governing equations are the Navier equations and the strain-displacement relations for the mechanical field, and the Gauss and Faraday laws for the electrostatic field. Moreover, the constitutive equations, assuming linear piezoelastic behavior, read:

$$D_i = d_{ikl}\sigma_{kl} + k_{ik}E_k \qquad (S1)$$

$$\epsilon_{ij} = s_{ijkl}\sigma_{kl} - d_{kij}E_k; \qquad (S2)$$

where $s_{ijkl}$, $d_{ikl}$, and $k_{ik}$ are the elastic compliance, piezoelectric, and dielectric permittivity coefficients, respectively, whereas $\epsilon_{ij}, \sigma_{ij}$ are the strain and stress components and $D_i, E_i$ are the electric displacement and the electric field components, respectively. The coupling between mechanical and electric fields is determined by the piezoelectric constants $d_{ikl}$. Eqs. S1 and S2 can be written in matrix form as:

$$\begin{bmatrix} \mathbf{D} \\ \boldsymbol{\epsilon} \end{bmatrix} = \begin{bmatrix} \mathbf{k}^\sigma & \mathbf{d}^d \\ -\mathbf{d}^c & \mathbf{s}^E \end{bmatrix} \begin{bmatrix} \mathbf{E} \\ \boldsymbol{\sigma} \end{bmatrix} \quad (S3)$$

where, for a three-dimensional solid, vectors **D** (C/m²) and **E** (V/m) have size (3×1), while vectors $\boldsymbol{\epsilon}$ (dimensionless) and $\boldsymbol{\sigma}$ (N/m²) have size (6×1). The latter two vectors are obtained from the corresponding second-order tensors using Voigt notation, e.g. the stress vector is:

$$\boldsymbol{\sigma} = \begin{bmatrix} \sigma_{11} \\ \sigma_{22} \\ \sigma_{33} \\ \sigma_{23} \\ \sigma_{31} \\ \sigma_{12} \end{bmatrix} \quad (S4)$$

In Eq. S3, the superscripts $c$ and $d$ allow one to distinguish the converse and the direct piezoelectric effects/coefficients. The superscripts $\boldsymbol{\sigma}$ and **E** indicate that the quantity is measured at constant stress and at constant electric field, respectively. The constant matrices are $\mathbf{k}^\sigma$ (F/m, 3×3 matrix), $\mathbf{d}^d$ (C/N or m/V, 3×6), $\mathbf{d}^c$ (C/N or m/V, 6×3), $\mathbf{s}^E$ (m²/N, 6×6). For a transversely isotropic material the piezoelectric, compliance and permittivity matrices take the following form:

$$\mathbf{d}^c = \begin{bmatrix} 0 & 0 & d_{31} \\ 0 & 0 & d_{32} \\ 0 & 0 & d_{33} \\ 0 & 0 & 0 \\ 0 & 0 & 0 \\ 0 & 0 & 0 \end{bmatrix} \quad (S5)$$

$$\mathbf{s}^E = \begin{bmatrix} s_{11} & s_{12} & s_{13} & 0 & 0 & 0 \\ s_{12} & s_{22} & s_{23} & 0 & 0 & 0 \\ s_{13} & s_{23} & s_{33} & 0 & 0 & 0 \\ 0 & 0 & 0 & s_{44} & 0 & 0 \\ 0 & 0 & 0 & 0 & s_{55} & 0 \\ 0 & 0 & 0 & 0 & 0 & s_{66} \end{bmatrix} \tag{S6}$$

$$\mathbf{k}^\sigma = \begin{bmatrix} k_{11} & 0 & 0 \\ 0 & k_{22} & 0 \\ 0 & 0 & k_{33} \end{bmatrix} \tag{S7}$$

Interestingly, since fibers are very long with respect to the cross-sectional diameter, plain strain effects (and thus a longitudinal stress) would arise even if the ends of the fibers were free to move longitudinally. The only difference would be that, in the latter case, two portions of the fibers close to the two ends would be free of longitudinal stress. The length of these portions, according to St. Venant's postulate,[S1] would be proportional to the maximum cross-sectional size and thus so small that the free end effect would be irrelevant for the global behavior.

## 3. Numerical model

To numerically model the cooperative behavior of nanostructures at the microscale, the fibers are discretized with linear 8-node brick elements featuring the piezoelastic constitutive behavior described above. The interaction between individual fibers is mainly due to contact between adjacent fibers as a result of the applied loading. For this reason, an important ingredient of the numerical model is the enforcement of frictionless electromechanical contact constraints at the interface between fibers. The implemented formulation is based on the classical master-slave concept. For each point on the slave surface, $x_s$, the corresponding point on the master surface is determined through normal (i.e. closest point) projection and is denoted as $x_m$. Thus the normal gap at each slave point is computed as:

$$g_N = (x_s - x_m) \cdot n \tag{S8}$$

$n$ being the outer unit normal to the master surface at the projection point. The sign of the measured gap is used to discriminate between active and inactive contact conditions, a negative value of the gap leading to active contact. The electric field requires the definition of the contact electric potential jump:

$$g_\phi = (\phi_s - \phi_m) \tag{S9}$$

where $\phi_s$ and $\phi_m$ are the electric potential values in the slave point and in its projection point on the master surface.

The discretization strategy used herein for the contact contribution is based on the node-to-surface approach combined with Bézier smoothing of the master surface. It is well known that the node-to-surface algorithm is susceptible of pathologies due to the $C^0$-continuity of the finite element (Lagrange) discretizations, and that these may affect the quality of results as well as iterative convergence in contact computations.[S2] One of the possible remedies are smoothing techniques for the master surface. Herein, the technique based on Bézier patches.[S3] and implemented within the AceGen/AceFEM environment is adopted and straightforwardly extended to electromechanical contact constraints.

The electromechanical constraints are regularized with the penalty method. The penalty parameters for the mechanical and for the electric contributions are appropriately chosen in order to obtain minimal penetration or voltage jump errors while avoiding ill-conditioning of the global stiffness matrix. According to standard finite element techniques, the global set of equations can be obtained by adding to the variation of the energy potential representing the continuum behavior the virtual work associated to the electromechanical contact contribution provided by the active contact elements. The final non-linear problem is consistently linearized with the automatic differentiation technique and solved using an incremental force-control procedure with adaptive time-stepping. If $\hat{\mathbf{u}}$ is the set of degrees of freedom (DOF) used to discretize the displacement field $\mathbf{u} = \mathbf{u}(\hat{\mathbf{u}})$, and $\hat{\boldsymbol{\phi}}$ is the set of DOF used to discretize the electric potential field $\varphi=\varphi(\hat{\boldsymbol{\phi}})$, so that $\hat{\mathbf{u}} \cup \hat{\boldsymbol{\phi}}$ is the vector of all nodal DOF, and $\Pi_{global}$ is the global energy of the discretized system, the residual vector and the stiffness matrix terms resulting from the finite element discretization are determined according to:

$$R_{u_i} = \frac{\delta \Pi_{global}}{\delta \hat{u}_i} \ ; \ R_{\phi_i} = \frac{\delta \Pi_{global}}{\delta \hat{\phi}_i} \ ;$$

$$K_{uu_{i,j}} = \frac{\delta R_{u_i}}{\delta \hat{u}_j} \ ; \ K_{\phi\phi_{i,j}} = \frac{\delta R_{\phi_i}}{\delta \hat{\phi}_j} \ ; \qquad (S10)$$

$$K_{u\phi_{i,j}} = \frac{\delta R_{u_i}}{\delta \hat{\phi}_j} \ ; \ K_{\phi u_{i,j}} = \frac{\delta R_{\phi_i}}{\delta \hat{u}_j} \ .$$

## 4. Further simulation results

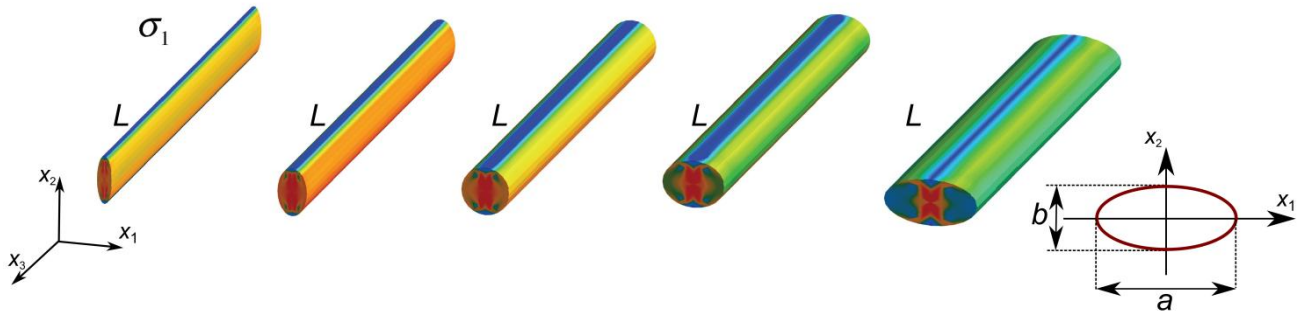

Fig. S2. Contour level of stress ($\sigma_1$) distribution in nanofibers with same length, *L*, and elliptical cross-section with different *b/a* ratios. Bottom inset: ellipse cross-section and *a* and *b* axes. Maximum and minimum stress values, corresponding to the used color scales (i.e. to red and blue, respectively) are, from left to right: +0.072 and − 0.296 MPa, +0.085 and − 0.314 MPa, +0.098 and − 0.368 MPa, +0.094 and − 0.348 MPa and +0.089 and − 0.246 MPa.

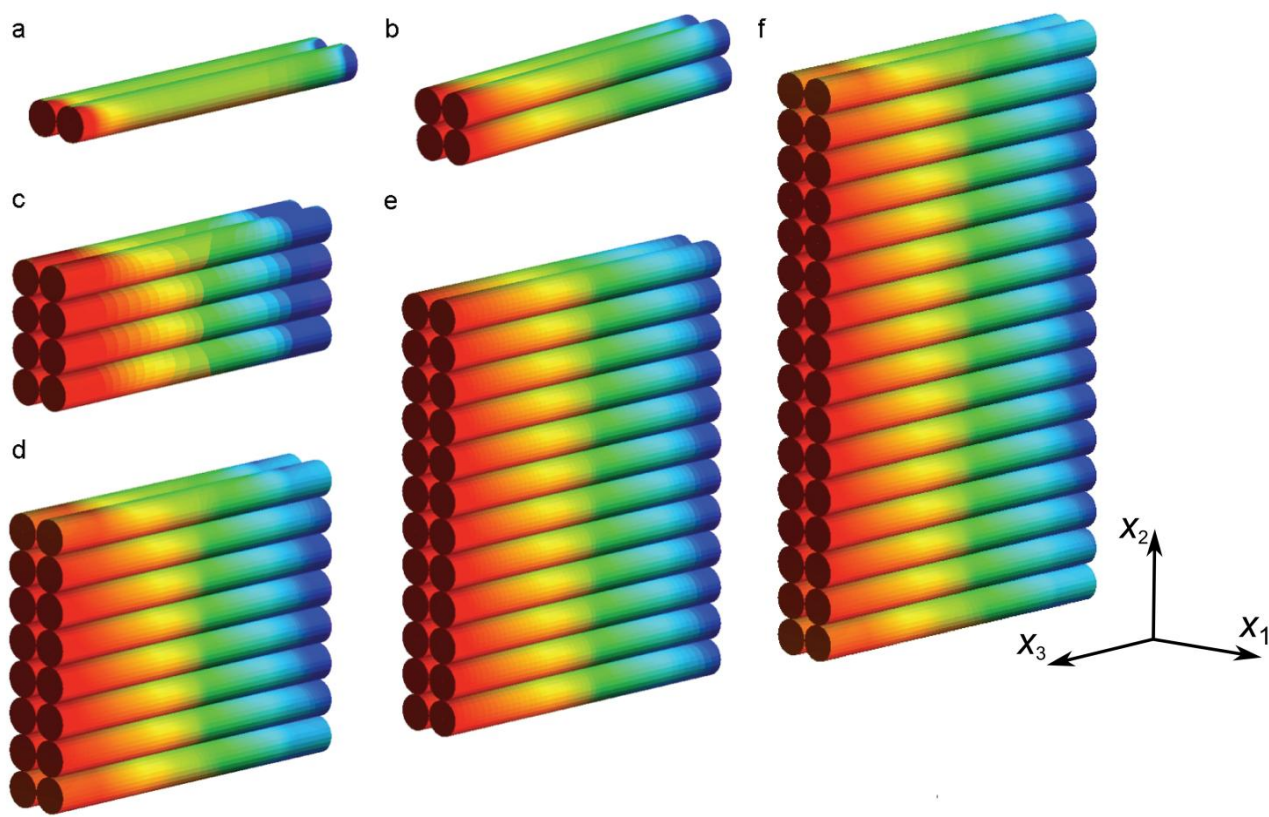

Fig. S3. Voltage distribution on the surface of nanofibers with circular cross-section and aligned in vertical arrays of increasing height. Maximum and minimum voltages, corresponding to the used color scales (i.e. to red and blue, respectively) are (a) ± 2.34 µV, (b) ± 10.94 µV, (c) ± 17.10 µV, (d) ± 27.95 µV, (e) ± 44.73 µV, and (f) ± 71.57 µV.

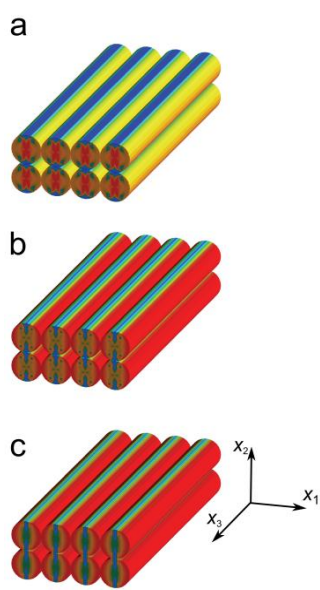

Fig. S4. Contour levels of stress distribution in the nanofibers, for the case of two overlapping layers. The components $\sigma_1$, $\sigma_2$, $\sigma_3$ are shown in (a), (b), and (c), respectively. Maximum and minimum stress values, corresponding to the used color scales (i.e. to red and blue, respectively) are, +0.198 and -0.863 MPa (a), +0.157 and − 0.749 MPa (b), +0.106 and − 0.461 MPa (c).

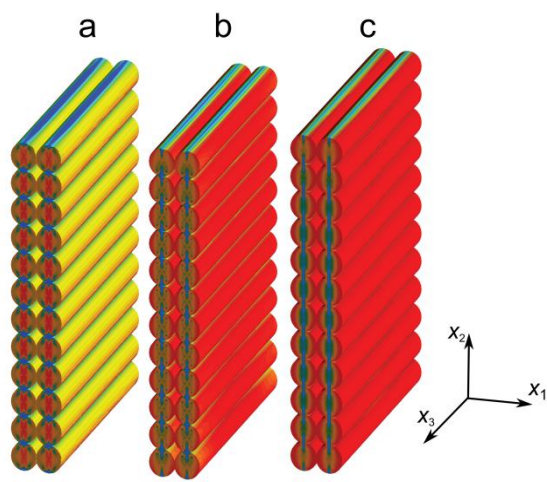

Fig. S5. Contour levels of stress distribution in the nanofibers, for the case of twelve overlapping layers. The components $\sigma_1$, $\sigma_2$ and $\sigma_3$ of the stress are shown in (a), (b), and (c) respectively. Maximum and minimum stress values, corresponding to the used color scales (i.e. to red and blue) are +0.415 and −1.231 MPa (a), +0.217 and −1.017 MPa (b), +0.363 and −0.962 MPa (c).

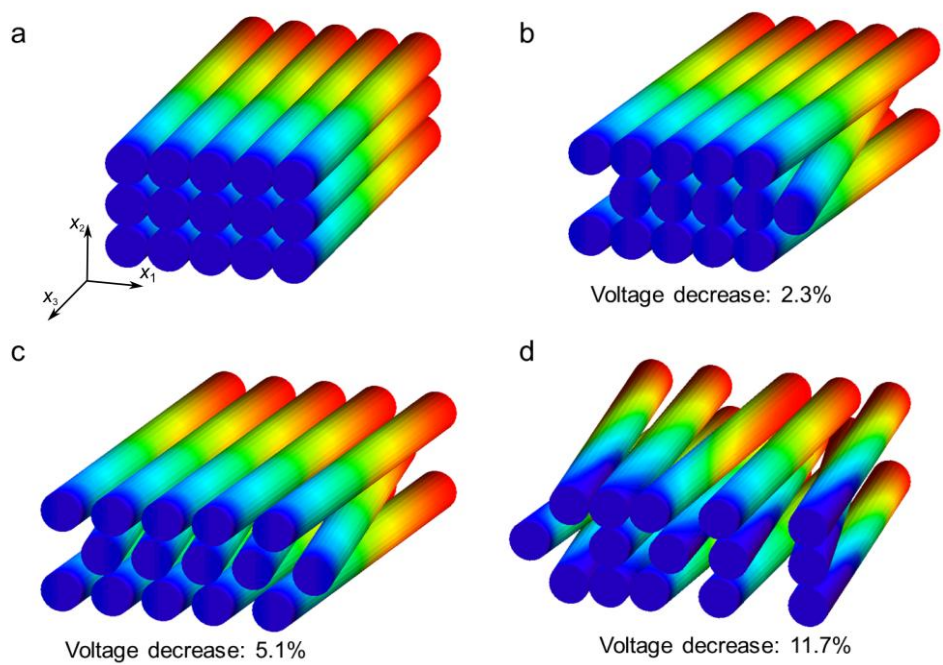

Fig. S6. Effect of fiber misalignment on the output voltage distribution. Three fiber layers stacked in the $x_2$ direction are subjected to an applied pressure = 120 Pa. (a): Perfect alignment (line contact) within each layer and among layers. (b): Misalignment of 10 degrees for all fibers in the intermediate layer (line contact within each layer, multipoint contact among layers). (c): Same as (b) but with mutual distance of one fiber radius between adjacent fibers (no contact within each layer, multipoint contact among layers). (d): Random fiber misalignment, angle range = ± 10 degrees, distance range = 0 to two fiber radii (multipoint contact within each layer and among layers). Maximum and minimum voltages, corresponding to the used color scales (i.e. to red and blue, respectively) are ± 18.624 μV (a), ± 18.190 μV (b), ± 17.670 μV (c), ± 16.440 μV (d).

## 5. Effective piezoelectric coefficients

Ideally, effective piezoelectric coefficients would be a more expressive measure of the piezoelectric performance for the system. However, we notice that the cooperative effect investigated in this paper is a *microstructural* effect. It arises from non-linear interactions taking place at the microscopic (i.e. at the fiber) scale, and is macroscopically observed as an effectively improved piezoelectric performance. Different macroscale geometries and boundary conditions may give rise to different microscale interactions, and therefore to a different effective macroscale performance. Therefore, a fully rigorous computation of effective piezoelectric coefficients 'once for all' is unfeasible. The most appropriate modeling framework to account for the microscale effects and concurrently determine the effective macroscale response is computational multiscale analysis,[S4] a recently emerged field of research in solid mechanics. An example is the so-called FE$^2$ approach, where a microscale boundary value problem with appropriate boundary conditions is solved for each quadrature point of the macroscale discretized geometry while solving a macroscale boundary value problem (within a finite element framework). This methodology is being followed by the authors in ongoing research but, due to its complexity, is outside of the scope of the present investigation.

A simpler but less rigorous approach to determine effective piezoelectric coefficients is computational homogenization. Following this strategy, effective coefficients are determined for a system with infinite fiber layers, each containing an infinite number of adjacent fibers. Due to the asymptotic behavior found in the analyses on the cooperative effect, the values found with this strategy correspond to the asymptotic ones.

**6. Homogenization procedure**

The main idea of homogenization is to find a globally homogeneous medium equivalent to the original heterogeneous one, where the equivalence is intended in an energetic sense as per Hill's balance condition. Coupling between the macroscopic and microscopic scales is here based on averaging theorems. Formulated for the electromechanical problem at hand, Hill's criterion in differential form reads:

$$\bar{\sigma}_{ij}\delta\bar{\epsilon}_{ij} + \bar{D}_i\delta\bar{E}_i = \frac{1}{V}\iiint \sigma_{ij}\delta\epsilon_{ij} dV + \frac{1}{V}\iiint D_i \delta E_i\, dV \qquad (S11)$$

and requires that the macroscopic volume average of the variation of work performed on the representative volume element (RVE) is equal to the local variation of work on the macroscale. In the previous equation: $\bar{\sigma}_{ij}$, $\bar{\epsilon}_{ij}$, $\bar{D}_i$ and $\bar{E}_i$ represent respectively the average values of stress, strain, electric displacement and electric field components while $V$ indicates the RVE volume. Hill's lemma leads to the following equations:

$$\bar{\sigma}_{ij} = \frac{1}{V}\iiint \sigma_{ij} dV; \quad \bar{\epsilon}_{ij} = \frac{1}{V}\iiint \epsilon_{ij} dV;$$
$$\bar{D}_i = \frac{1}{V}\iiint D_i dV; \quad \bar{E}_i = \frac{1}{V}\iiint E_i dV \qquad (S12)$$

Classically three types of boundary conditions are used for an RVE: prescribed displacements, prescribed tractions and periodic boundary conditions. In this work, we used periodic boundary conditions. This implies that the RVE represents a periodic structure and thus the system consists of infinite fiber layers, each containing an infinite number of adjacent fibers. The periodic boundary conditions expressed as linear constraints are implemented in AceGen as multipoint constraints using the Lagrange multiplier method. For simplicity, meshing of the RVE is performed uniformly such that identical nodes are present on all faces of the RVE. The final constitutive equation in the homogenized setting reads:

$$\begin{pmatrix}\bar{D}\\ \bar{\epsilon}\end{pmatrix} = \begin{pmatrix}\bar{k} & \bar{d}\\ -\bar{d} & \bar{s}\end{pmatrix}\begin{pmatrix}\bar{\sigma}\\ \bar{E}\end{pmatrix} \qquad (S13)$$

where $\bar{s}_{ijkl}$, $\bar{d}_{ikl}$, and $\bar{k}_{ik}$ are respectively the homogenized compliance, piezoelectric strain, and dielectric constants.